\documentclass[prc,twocolumn,showpacs,preprintnumbers]{revtex4}
\usepackage{bm}
\usepackage{amssymb}
\usepackage{amsmath}
\usepackage{graphicx}
\begin{document}
%%%%%%%%%%%
%%%%%%%%%%%
\title{Fusion of neutron rich oxygen isotopes in the crust of accreting neutron stars}
%%%
\author{C. J. Horowitz}\email{horowit@indiana.edu} 
\author{H. Dussan}
\affiliation{Department of Physics and Nuclear Theory Center,
             Indiana University, Bloomington, IN 47405}
\author{D. K. Berry}\email{dkberry@indiana.edu}
\affiliation{University Information Technology Services,
             Indiana University, Bloomington, IN 47408}

%%%%%%%%%%%
\date{\today}
\begin{abstract}
Fusion reactions in the crust of an accreting neutron star are an important source of heat, and the depth at which these reactions occur is important for determining the temperature profile of the star.  Fusion reactions depend strongly on the nuclear charge $Z$.  Nuclei with $Z\le 6$ can fuse at low densities in a liquid ocean.  However, nuclei with $Z=8$ or 10 may not burn until higher densities where the crust is solid and electron capture has made the nuclei neutron rich.  We calculate the $S$ factor for fusion reactions of neutron rich nuclei including $^{24}$O + $^{24}$O and $^{28}$Ne + $^{28}$Ne.  We use a simple barrier penetration model.  The $S$ factor could be further enhanced by dynamical effects involving the neutron rich skin.  This possible enhancement in $S$ should be studied in the laboratory with neutron rich radioactive beams.  We model the structure of the crust with molecular dynamics simulations.  We find that the crust of accreting neutron stars may contain micro-crystals or regions of phase separation.  Nevertheless, the screening factors that we determine for the enhancement of the rate of thermonuclear reactions are insensitive to these features.  Finally, we calculate the rate of thermonuclear $^{24}$O + $^{24}$O fusion and find that $^{24}$O should burn at densities near $10^{11}$ g/cm$^3$.  The energy released from this and similar reactions may be important for the temperature profile of the star.  
\end{abstract}
\smallskip
\maketitle

\section{Introduction}
Nuclei accreting onto a neutron star undergo a variety of reactions.  First at low densities, conventional thermonuclear fusion takes place, see for example \cite{rpash}.  Next as nuclei are buried to higher densities, the rising electron Fermi energy induces a series of electron captures \cite{gupta}.  Finally at very high densities, nuclei can fuse via pycnonuclear reactions.   These reactions are induced by the quantum zero point motion \cite{pycno}.   The energy released, and the densities at which reactions occur, are important for determining the temperature profile of neutron star crusts.

Superbursts are very energetic X-ray bursts from accreting neutron stars that are thought to involve the unstable thermonuclear burning of carbon \cite{superbursts, superbursts2}.  However, some simulations do not reproduce the conditions needed for carbon ignition because they have too low temperatures \cite{superignition}.  An additional heat source, from fusion or other reactions, could raise the temperature and allow carbon ignition at densities that reproduce observed burst frequencies.

Recently the cooling of two neutron stars has been observed after extended outbursts \cite{Wijnands, cackett}.  These outbursts heated the crusts out of equilibrium and then the cooling time was measured as the crusts returned to equilibrium.  The surface temperature of the neutron star in KS 1731-260 decreased with an exponential time scale of 325 $\pm$ 100 days while MXB 1659-29 has a time scale of 505 $\pm$ 59 days \cite{cackett}.  These cooling times depend on the thermal conductivity of the crust and the initial temperature profile.  Comparing these observations of relatively rapid cooling to calculations by Rutledge et al. \cite{rutledge} and Shternin et al. \cite{shternin} suggests that the crust has a high thermal conductivity.  However, if the initial temperature profile of the crust is peaked near the surface, then this peak could quickly diffuse to the surface and lead to rapid cooling.  Therefore, cooling time scales are also sensitive to the initial temperature profile, and this depends on heating from nuclear reactions at moderate densities in the crust.

Gupta et al. have calculated heating from electron capture reactions in the outer crust \cite{gupta}.  While they find more heating than previous works, they still find no more than 0.4 MeV per nucleon total heating from all of the electron captures on any mass number $A$ system.  Haensel and Zdunik have calculated pycnonuclear fusion reactions at great densities in the inner crust \cite{haensel}.  However, if reactions occur deep in the inner crust, most of the heat may flow in to the core instead of out towards the surface.  As a result, there may be a smaller impact on the temperature profile of the outer crust.         

A low crust thermal conductivity, for example from an amorphous solid, could help explain superburst ignition.  This could better insulate the outer crust and allow higher carbon ignition temperatures.  However, a low thermal conductivity appears to be directly contradicted by the observed short crust cooling times.  Furthermore, our molecular dynamics simulations in ref. \cite{horowitz} and further results we present in Section \ref{MD} find a regular crystal structure, even when the system has a complex composition with many impurities.  We do not find an amorphous phase.  These results will be discussed further in a later publication.  We conclude that the thermal conductivity of the crust is high.

If the thermal conductivity is high, one may need additional heat sources, at moderate densities, in order to explain superburst ignition.  Although Gupta et al. find additional heating from electron captures to excited nuclear states, simple nuclear structure properties may provide a natural limit to the total heating from electron captures \cite{brownprivate}.  Haensel and Zdunik \cite{haensel,haensel2007} consider heating from pycnonuclear reactions using a simple one component plasma model.  They find that fusion reactions may not take place until relatively high densities above $10^{12}$ g/cm$^3$.  However, their use of a one component plasma could be a significant limitation.  Fusion reactions depend strongly on the nuclear charge $Z$.  Therefore, the reaction rate may be highest for the rare impurities that have the lowest $Z$, instead of for nuclei of average charge.  

In this paper, we go beyond Haensel and Zdunik and consider a full mixture of complex composition instead of assuming one average charge and mass.  We focus on thermonuclear and pycnonuclear reactions at densities around $10^{11}$ g/cm$^3$.  This is near the base of the outer crust.  Heat released at this density could be important for superburst ignition and for crust cooling times.  Nuclei at this density are expected to be neutron rich.  Furthermore, the other nearby ions strongly screen the Coulomb barrier and greatly enhance the rate of thermonuclear reactions. 

We begin by describing the initial composition.  This includes neutron rich light nuclei such as $^{24}$O and $^{28}$Ne.  We calculate cross sections and $S$ factors for $^{24}$O + $^{24}$O and $^{28}$Ne + $^{28}$Ne fusion using a simple barrier penetration model.  Note that the dynamics of the neutron rich skins of these nuclei can enhance the cross section over that predicted by our simple barrier penetration model.  This is a very interesting and open nuclear structure question, see for example \cite{subbarrier}.  

Next, we use classical molecular dynamics simulations to determine the structure of the crust and screening factors for the enhancement of thermonuclear reactions.  There are many previous calculations of screening factors for the one component plasma \cite{ocpscreening} and for binary ion mixtures, see for example \cite{bimscreening}.  However, we are not aware of any previous calculations for a crystal of a complex multicomponent composition.   Finally, we calculate reaction rates and conclude that $^{24}$O is expected to fuse at densities near $10^{11}$ g/cm$^3$ while $^{28}$Ne should react at densities near $10^{12}$ g/cm$^3$.  Heat from these reactions may be important for determining the temperature profile of accreting neutron stars.

\section{Crust Composition}
\label{composition}

We now describe our model for the composition of the crust.  This is the same as was used in previous work on chemical separation when the crust freezes \cite{horowitz}.  Schatz et al. have calculated the rapid proton capture (rp) process of hydrogen burning on the surface of an accreting neutron star \cite{rpash}, see also \cite{rpash2}.  This produces a variety of nuclei up to mass $A\approx 100$.  Gupta et al. then calculate how the composition of this rp process ash evolves, because of electron capture and light particle reactions, as the material is buried by further accretion.  Their final composition, at a density of $2.16\times 10^{11}$ g/cm$^3$, has forty \% of the ions with atomic number $Z=34$, while an additional 10\% have $Z=33$.  The remaining 50\% have a range of lower $Z$ from 8 to 32.  In particular about 3\% is $^{24}$O and 1\% $^{28}$Ne.  This Gupta et al. composition is listed in Table \ref{tablezero}.  In general, nuclei at this depth in the crust are expected to be neutron rich because of electron capture.

\begin{table}
\caption{Abundance $x_i$ (by number) of chemical element $Z$ and average mass number $\langle A\rangle$. } 
\begin{tabular}{lll}
$Z$ & Abundance ($x_i$) & $\langle A\rangle$ \\
8 & 0.0301  & 24\\
10 & 0.0116 & 28.8 \\
12 & 0.0023 & 36\\
14 & 0.0023 & 42\\
15 & 0.0023 & 45\\
20 & 0.0046 & 62\\
22 & 0.0810 & 66.06 \\
24 & 0.0718 & 74\\
26 & 0.1019 & 76\\
27 & 0.0023 & 77\\
28 & 0.0764 & 80\\
30 & 0.0856 & 89.35\\
32 & 0.0116 & 96\\
33 & 0.1250 & 99\\
34 & 0.3866 & 102.61\\
36 & 0.0023 & 106\\
47 & 0.0023 & 109\\
\end{tabular} 
\label{tablezero}
\end{table}

\section{Fusion Cross Sections and $S$ Factors}
\label{crosssections}

There is a great deal of experimental information on low energy fusion cross sections for light stable nuclei such as $^{12}$C  \cite{carbon, gasques} and $^{16}$O \cite{oxygen}.  For these nuclei, barrier penetration models work well \cite{bp}.  However, recently Jiang et al.  discuss fusion hindrance at extreme sub coulomb barrier energies \cite{jiang}.  Much less information is available for the fusion of very neutron rich light nuclei.  We use a simple barrier penetration model to calculate fusion cross sections for $^{24}$O and $^{28}$Ne.  We start with the Sao Paulo double folding potential $V_F(r)$ \cite{sp},
\begin{equation}
V_F(r)=\int d^3r_1d^3r_2 \rho_1(r_1)\rho_2(r_2) V_0 \delta({\bf r}_1- {\bf r_2} - {\bf r})\, .
\label{vf}
\end{equation}   
Here $\rho_1$ and $\rho_2$ are the densities of the two nuclei and $V_0=-456$ MeV-fm$^{3}$.  Next tunneling through the Coulomb barrier is calculated in a WKB approximation including some nonlocality effects \cite{sp, pycno}.  For simplicity we assume Wood Saxon densities with radius parameter $R=1.31A^{1/3}-0.84$ fm and diffuseness $a=0.58$ fm \cite{sp,pycno}.  These parameters reproduce the measured cross sections for $^{16}$O+$^{16}$O fusion.  Our results for the fusion cross section $\sigma(E)$ at center of mass energy $E$ are expressed as the astrophysical $S$ factor, $S=E\sigma(E)e^{2\pi\eta}$ and collected in Table \ref{tableone}.  Here the Gamow penetration factor is $\eta=Z_1Z_2e^2(\mu/2E)^{1/2}$, the nuclei have charges $Z_1$ and $Z_2$, and $\mu$ is the reduced mass.  Our $S$ factor for $^{24}$O+$^{24}$O is over eight orders of magnitude larger than that for $^{16}$O+$^{16}$O.  We have also calculated $S$ for $^{24}$O+$^{24}$O using relativistic mean field densities calculated with the NL3 interaction \cite{NL3}.  This yields $S$ that is only slightly higher than the calculation with Wood Saxon densities.  In addition, Gasques et al. \cite{gasques07} have calculated $S$ for $^{24}$O+$^{24}$O using both a fermionic molecular dynamics model and the Sao Paulo model of Eq. \ref{vf}, and they obtain similar results.  

\begin{table}
\caption{Astrophysical $S$ factors for low energy fusion reactions versus center of mass energy $E$.} 
\begin{tabular}{llll}
$E$ & $^{24}$O+$^{24}$O & $^{28}$Ne+$^{28}$Ne \\
\toprule
(MeV) & (MeV-barn) & (MeV-barn) \\
1 & $1.7\times 10^{35}$ & \\
1.5 & $1.1\times 10^{35}$ & $1.0 \times 10^{48}$ \\
2 & $7.4\times 10^{34}$ & $7.2 \times 10^{47}$ \\
3 & $3.2 \times 10^{34}$ & $3.4 \times 10^{47}$ \\
4 & $1.3\times 10^{34}$ & $1.6 \times 10^{47}$ \\
6 & $1.9\times 10^{33}$ & $3.4 \times 10^{46}$ \\
8 & $2.5\times 10^{32}$ & $6.4 \times 10^{45}$ \\
\end{tabular} 
\label{tableone}
\end{table}

These barrier penetration results may provide lower limits for the cross sections.  Dynamical effects, not included in Eq. \ref{vf}, can increase the cross section.  Indeed, low energy cross sections are observed to be larger than simple barrier penetration estimates for heavier stable nuclei \cite{enhancement}.   The extended neutron skin of very neutron rich nuclei presents a very interesting special case for low energy fusion reactions.  The dynamics of the easily polarizable skin can increase the cross section.  For example, a neutron rich neck could form between the nuclei decreasing the Coulomb barrier.  The dynamics of the skin, and its effects on low energy fusion,  are very important nuclear structure questions that should be studied further with radioactive beams.  For example, it should be possible to measure low energy fusion cross sections for beams of a neutron rich O isotope colliding with a stable light target such as $^{16}$O or $^{12}$C.

\section{Molecular Dynamics Simulations of Crust Structure}
\label{MD}

We now consider fusion reactions in a dense medium.  There have been many calculations of the strong screening enhancement of thermonuclear reactions and of pycnonuclear reactions.  For example, recently Gasques et al. \cite{gasques} presented a phenomenological formula for reactions in a one component plasma (OCP) that is valid for all regimes of density and temperature.  This has been extended by Yakovlev et al. to multicomponent plasmas (MCP) \cite{yakovlev}.  

In order to calculate pycnonuclear reactions in a multicomponent system one needs to understand its state.  Monte Carlo simulations \cite{mcocp} of the freezing of a classical OCP indicate that it can freeze into imperfect body centered cubic (bcc) or face-centered cubic (fcc) microcrystals.  Unfortunetly not much has been published on the freezing of MCP.  In an earlier work \cite{horowitz}, we calculated chemical separation upon freezing of our MCP composed of rp process ash.  We found, based on large scale molecular dynamics (MD) simulations, that chemical separation takes place.  The liquid phase is greatly enriched in low $Z$ elements compared to the solid phase.  We found that the solid phase formed a regular lattice where the charge of a given lattice site was more or less random.  However, we did not study the structure of the solid in detail, aside from its average composition.       

There are many possibilities for the state of a cold MCP \cite{yakovlev}.  It can be a regular MCP lattice; or microcrystals; or an amorphous, uniformly mixed structure; or a lattice of one phase with random admixture of other ions; or even an ensemble of phase separated domains.  We perform classical MD simulations to explore the state of our MCP solid.  The electrons form a very degenerate relativistic electron gas that slightly screens the interaction between ions.  We assume the potential $v_{ij}(r)$ between the ith and jth ion is,
\begin{equation}
v_{ij}(r) = \frac{Z_i Z_j e^2}{ r} {\rm e}^{-r/\lambda_e}\, ,
\label{vij}
\end{equation}
where $r$ is the distance between ions and the electron screening length is $\lambda_e=\pi^{1/2}/2e(3\pi^2 n_e)^{1/3}$.  Here $n_e$ is the electron density.

To characterize our simulations , we define Coulomb coupling parameters $\Gamma_j$ for ions of charge $Z_j$,
\begin{equation}
\Gamma_j=\frac{Z_j^2e^2}{a_j T}\, ,
\label{gammaj}
\end{equation}
with $T$ the temperature and $a_j$ is the radius of a sphere containing $Z_j$ electrons (the ion sphere radius),
\begin{equation}
a_j=Z_j^{1/3} \Bigl(\frac{3}{4\pi n_e}\Bigr)^{1/3}\, .
\label{aj}
\end{equation}
The average coupling parameter $\Gamma$ for the MCP is,
\begin{equation}
\Gamma=\sum_j \Gamma_j x_j \ \ \ = \frac{\langle Z^{5/3} \rangle \langle Z \rangle^{1/3} e^2}{a T}\, ,
\label{gamma}
\end{equation}
where $x_j$ is the abundance (by number) of charge $Z_j$ and the overall ion sphere radius $a$ is $a=(3/4\pi n)^{1/3}$ with $n=n_e/\langle Z\rangle$ the ion density.   The OCP freezes at $\Gamma=175$.  In ref. \cite{horowitz} we found that the impurities in our MCP lowered the melting temperature until $\Gamma=247$.  Finally, we can measure time in our simulation in units of one over the plasma frequency $\omega_p$,
\begin{equation}
\omega_p=\Bigl(\sum_j\frac{Z_j^2 4\pi e^2 x_i n}{M_i}\Bigr)^{1/2}\, ,
\label{omegap}
\end{equation}
where $M_j$ is the average mass of ions with charge $Z_j$.

To explore possible states for the multicomponent plasma we perform two molecular dynamics simulations.  The initial conditions of these simulations are similar to those in \cite{horowitz}.  We start by freezing a very small system of 432 ions.  Here the ions were started with random initial conditions at a high temperature $T$ and $T$ was reduced in stages (by re-scaling velocities) until the system freezes.  For the first simulation run, called rpcrust-01 in Table \ref{tabletwo}, we place four copies of this 432 ion solid in a larger simulation volume along with four copies of a 432 ion liquid configuration.  This 3456 ion configuration is evolved at a lower temperature until the whole system freezes.  Finally, we evolve the 3456 ion solid at a reference high density of $n=7.18\times 10^{-5}$ fm$^{-3}$ (or $1\times 10^{13}$ g/cm$^3$) and a temperature of $T=0.325$ MeV for a total simulation time of $2.4\times 10^9$ fm/c ($8.9\times 10^6$ $\omega_p^{-1}$).   This density and temperature correspond to $\Gamma=261.6$.  Evolution was done using the velocity verlet algorithm \cite{verlet} using a time step of $\Delta t=25$ fm/c for a total of $9.6\times 10^7$ steps.  This took about 2 months on a single special purpose MDGRAPE-2 \cite{mdgrape} board.  The simulation results can be scaled to other densities $n'$ and temperatures $T'$ that also correspond to $\Gamma=261.6$.  IE $n'^{1/3}/T'=(7.18\times 10^{-5} {\rm fm}^{-3})^{1/3}/(0.325 {\rm MeV})$.  

\begin{table}
\caption{Simulation Parameters.  The temperature $T$, Coulomb parameter $\Gamma$, total simulation time $t$, average charge $\langle Z\rangle$, and average $Z^{5/3}$ for the two simulations.  Each simulation is at a density $n=7.18\times 10^{-5}$ fm$^{-3}$ ($1\times 10^{13}$ g/cm$^3$) and involves 3456 ions.} 
\begin{tabular}{llllll}
Run & $T$(MeV) & $\Gamma$ & t(fm/c) & $\langle Z \rangle$ & $\langle Z^{5/3} \rangle$ \\
\toprule
rpcrust-01 & 0.325 & 261.6 & $2.4\times 10^9$ & 29.3 & 285.8 \\
rpcrust-02 & 0.35  & 242.9 & $1.6\times 10^9$ & 29.3 & 285.8 \\
\end{tabular} 
\label{tabletwo}
\end{table}

\begin{figure}[ht]
\begin{center}
\includegraphics[width=3in,angle=0,clip=true] {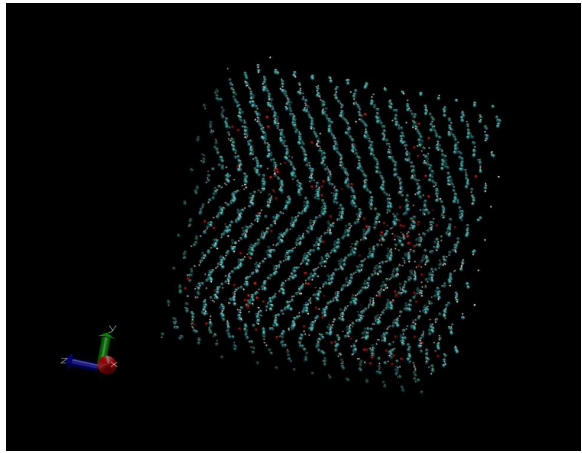}
\caption{(Color on line) Configuration of the 3456 ion mixture in run rpcrust-01 after a simulation time of $2.4 \times 10^9$ fm/c at $\Gamma=261.6$.  The small red spheres show the positions of $^{24}$O ions, while ions of above average $Z$ are shown as larger blue spheres.  Finally, ions of below average $Z$ (except for O) are shown as small white spheres. The upper and lower halves of the figure show two micro-crystals of different orientations.}
\label{Fig1}
\end{center}
\end{figure}

The final configuration  for run rpcrust-01, see Table \ref{tabletwo}, is shown in Fig. \ref{Fig1} after a simulation time of $2.4\times 10^9$ fm/c.  The system is seen to be composed of two micro-crystals of different orientations.  This is similar to the micro-crystals found in ref. \cite{mcocp} upon freezing a one component plasma.  In Fig. \ref{Fig1} we highlight the positions of the $^{24}$O ions (as small red spheres).  These ions are located both in the crystal planes and in between them.  The O ions are not spread uniformly throughout the volume but there is a tendency for them to cluster.  This will be discussed in more detail below.

\begin{figure}[ht]
\begin{center}
\includegraphics[width=3in,angle=0,clip=true] {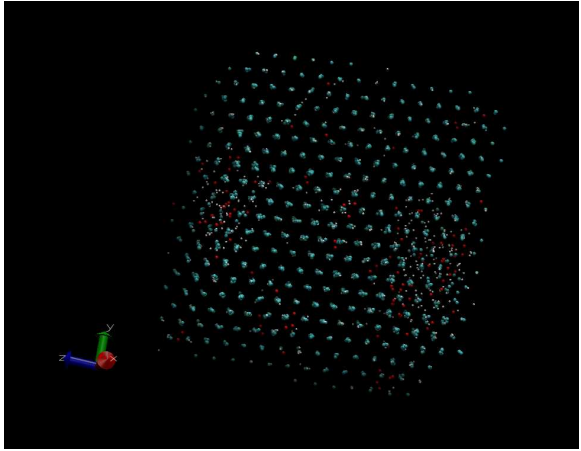}
\caption{(Color on line) Configuration of the 3456 ion mixture in run rpcrust-02 after a simulation time of $1.6 \times 10^9$ fm/c at $\Gamma=242.9$.  The small red spheres show the positions of $^{24}$O ions, while ions of above average $Z$ are shown as larger blue spheres.  Finally, ions of below average $Z$ (except for O) are shown as small white spheres.  The $^{24}$O concentration is seen to be enhanced in sub-regions.}
\label{Fig2}
\end{center}
\end{figure}

Given the micro-crytals for run rpcrust-01, we performed a second simulation, labled rpcrust-02 in Table \ref{tabletwo}, with different initial conditions.  Here eight copies of a 432 ion solid configuration were placed in the larger simulation volume.  The system was evolved at a slightly higher temperature $T=0.35$ MeV to possibly speed the diffusion of O ions.  The total simulation time is $1.6\times 10^9$ fm/c.  Note, that this run is ongoing and results for longer simulation times will be reported in a later publication.  Figure \ref{Fig2} shows the final configuration of the 3456 ions.  Now the system involves a single body-centered cubic (bcc) crystal.  The O ions are not uniformly distributed.  Instead O is strongly enriched in local regions.  This simulation has $\Gamma=243$.  This corresponds to a significantly lower temperature than the melting temperature of a pure OCP which corresponds to $\Gamma\approx 175$.  However, the impurities were found in ref. \cite{horowitz} to lower the melting temperature till $\Gamma\approx 247$.  Therefore the O rich regions in rpcrust-02 may be related to the formation of a bulk liquid phase which was found in ref. \cite{horowitz} to be greatly enriched in O.

We now use these simulation results to calculate the effective screening potential $v_{eff}^i(r)$ provided by all of the other ions.  This greatly enhances the rate of thermonuclear fusion reactions of two charge $Z_i$ ions.
\begin{equation}
v_{eff}^i(r)=-T\, {\rm ln} g_{ii}(r) - \frac{Z_i^2e^2}{r}
\label{veff}
\end{equation}
The radial distribution function $g_{ii}(r)$ gives the probability to find another ion of charge $Z_i$ a distance $r$ away from a given charge $Z_i$ ion.  This is normalized to one at large distances $g_{ii}(r\rightarrow \infty)\rightarrow 1$.  We calculate $g_{ii}$ from our simulation by histograming relative distances.  Figure \ref{Fig2} shows $g_{ii}(r)$ for $Z_i=8$ (O), 10 (Ne), 22 (Ti), and 34 (Se).  Note that because of the low Ne abundance, our $g_{ii}$ results for $Z_i=10$ are based on the positions of only 40 ions!  Therefore we caution that our Ne results may have large finite size and or statistical errors.

Selenium is the dominate species.  Therefore the Se ion locations largely determine the bulk structure of the crystal lattice.  The distance between peaks in $g_{ii}$ for Se reflects the lattice spacing.  The radial distribution function for Ti closely follows that for Se at large distances.  This shows that Ti, for the most part, occupies the same lattice sites as Se.  However the first peak in $g_{ii}$ for Ti occurs at smaller distances than the first peak for Se.  This reflects the smaller ion sphere radius $a_j$ Eq. \ref{aj} for the lower charged Ti because the coulomb repulsion between two Ti ions is smaller than that between two Se ions.  

The radial distribution functions for O and Ne show large peaks at small distances.  This may reflect a tendency to replace a single Se ion with a cluster of multiple low charge O or Ne ions.  The radial distribution function for O does not have large dips between the peaks in $g_{ii}$ for Se.  This shows that O also occupies positions in between the lattice planes.  Finally, $g_{ii}$ for O and Ne is larger than one at intermediate and large distances.  This shows that the the low $Z$ ions are not uniformly distributed.  Instead they preferentially cluster in sub regions that are greatly enriched in low $Z$ ions.         

\begin{figure}[ht]
\begin{center}
\includegraphics[width=2.8in,angle=270,clip=true] {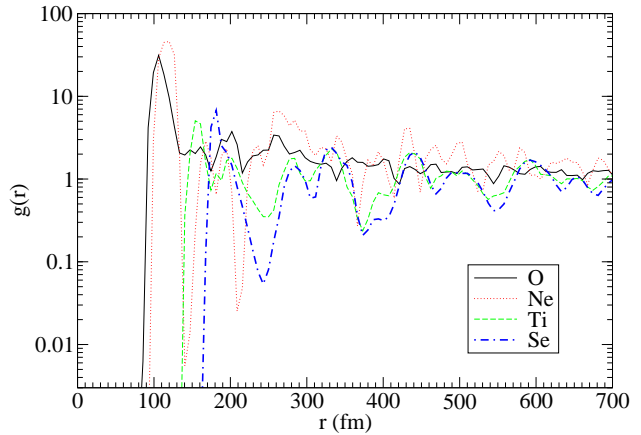}
\caption{(Color on line) Radial distribution functions $g_{ii}(r)$ for ions of charges $Z_i=8$(O), 10(Ne), 22(Ti), and 34 (Se).  Results from run rpcrust-01 at a simulation time of $2.4 \times 10^9$ fm/c have been scaled to a density of $3.6 \times 10^{10}$ g/cm$^3$ and $T=0.05$ MeV. }
\label{Fig3}
\end{center}
\end{figure}
 
\begin{figure}[ht]
\begin{center}
\includegraphics[width=2.8in,angle=270,clip=true] {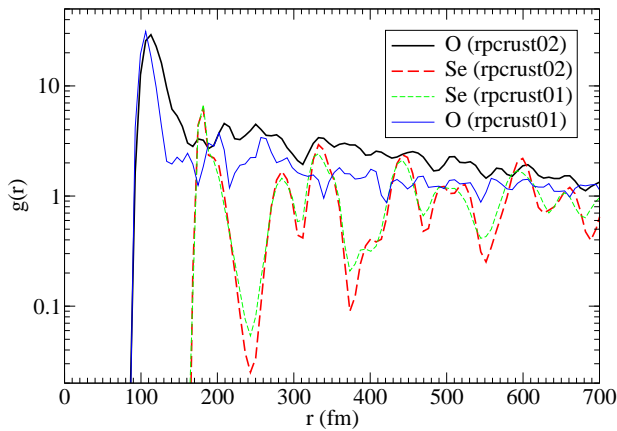}
\caption{(Color on line) Radial distribution functions $g_{ii}(r)$ for ions of charges $Z_i=8$ (solid lines) and 34 (dashed lines) at a density of $3.6 \times 10^{10}$ g/cm$^3$.  Heavy lines show run rpcrust-02 results at a simulation time of $1.6\times 10^{9}$ fm/c while thin lines show run rpcrust-01 results at $t=2.4 \times 10^9$ fm/c. }
\label{Fig4}
\end{center}
\end{figure}

The radial distribution functions for run rpcrust-02 are compared to run rpcrust-01 in Fig. \ref{Fig4}.  Results for Se are similar.  However, $g_{ii}(r)$ for O, at intermediate and large distances, is even more enhanced for rpcrust-02 than for rpcrust-01.  Presumably, the slightly higher temperature of rpcrust-02 enhances phase separation into regions that are enriched in low $Z$ ions.  
 
We now use these $g_{ii}$ results to calculate reaction rates.  Strong ion screening enhances the rate of thermonuclear fusion by a factor $F$, see for example \cite{yakovlev},
\begin{equation}
F={\rm exp}(h^j(r=0))\, .
\label{F}
\end{equation}
Here $h^j(r)=-v_{eff}^j(r)/T$.  For simplicity, we neglect the dependence of $F$ on $h^j(r)$ for $r\ne 0$, see \cite{jancovici}.  Unfortunately, it is hard to get good statistics on $g_{ii}(r)$ for small $r$.  Therefore, one must extrapolate our MD results for $r\approx a_j$ to smaller $r$.  The form of $h(r)$ for small $r$ is known for the OCP.  We generalize the OCP expression for $h(r)$ in ref. \cite{OCPh} to the multi-component case and assume for $r\leq 1.5 a_j$,
\begin{equation}
h^j(r)\approx h^j_0 -\frac{1}{4} \Gamma_j \Bigl(\frac{r}{a_j}\Bigr)^2 + 0.0277 \Gamma_j \Bigl(\frac{r}{a_j}\Bigr)^4\, .
\label{happrox}
\end{equation}
We fit Eq. \ref{happrox} to our MD results for $g_{ii}(r)$ over the range where we find a nonzero $g_{ii}(r)$ and $r \leq 1.5 a_j$ and extract values for $h_0^j=h^j(r=0)$.  The enhancement of the thermonuclear rate is then $F={\rm Exp}(h^j_0)$.  Our values for $h_0^j$ are collected in Table \ref{tablethree}.    These values are averages of five values of $h_0^j$ calculated at t=2.0, 2.1, 2.2, 2.3, and  $2.4\times 10^9$ fm/c for run rpcrust-01 and at t=1.2, 1.3, 1.4, 1.5, and $1.6\times 10^9$ fm/c for run rpcrust-02.  At each time, we calculate $h_0^j$ from the average of 25000 configurations separated by 250 fm/c.  A simple analytic formula for $h_0^j$, based on the linear mixing rule for the MCP is, see for example \cite{yakovlev},
\begin{equation}
h^j(r=0)\approx 1.0754 \Gamma_j.
\label{h0analytic}
\end{equation}
We find good agreement between our MD simulation results and Eq. (\ref{h0analytic}) even at our large value of $\Gamma=262$.  Furthermore this is true for both rpcrust-01 and rpcrust-02 runs.  For the large $Z$ ions Ti and Se, the agreement between our MD results and Eq. (\ref{h0analytic}) is better than 1\%.  For the lighter ions O and Ne, our MD results are slightly larger than Eq. (\ref{h0analytic}).  Again we caution that our MD results for Ne are based on only 40 ions.  Therefore we focus on O.  Our enhancement for O may reflect the clustering of O into sub-regions as shown in Fig. \ref{Fig2}.

In any case, the overall agreement between our MD results and Eq. (\ref{h0analytic}) is very good.  It is a major result of this paper that features in the crust, such as the micro-crystals in Fig. \ref{Fig1}, or the phase separation in Fig. \ref{Fig2}, do not appear to be important for the screening of thermonuclear reactions.  Therefore, for the rest of this paper we will simply use Eq. (\ref{h0analytic}) to describe screening.  Note that these features in the crust may be much more important for calculating pycno-nuclear reaction rates at very high densities.  This will be explored in later work.

\begin{table}
\caption{Screening potential $h^j(r=0)$ for O, Ne, Ti, and Se ions ($Z_j=8,10,22, 34$) from our MD simulations.  Also listed is an analytic approximation $h^j(r=0)=1.0754\Gamma_j$, Eq. (\ref{h0analytic}) and the difference between our MD results and the analytic approximation.} 
\begin{tabular}{llllll}
Run & $\Gamma$ & O & Ne & Ti & Se \\
\toprule
rpcrust-01 &  261.6 & 32.5 & 48.6 & 169.6 & 350.0  \\
analytic & 261.6 & 31.5 & 45.7 & 170.0 & 351.2 \\
MD\, -\, analytic &261.6  & 1.0 & 2.9 & -0.4 & -1.2 \\
rpcrust-02 & 242.9 & 30.3 & 43.8 & 158.1 & 325.5 \\
analytic & 242.9 & 29.2 & 42.4 & 157.8 & 326.1 \\
MD\, -\, analytic & 242.9  & 1.1 & 1.4 & 0.3 & -0.6 \\
\end{tabular} 
\label{tablethree}
\end{table}

\section{Reaction Rates}
\label{reaction rates}

We now calculate the rate of thermonuclear $^{24}$O+$^{24}$O fusion including the effects of strong screening.  The reaction rate per O ion $R/n$ is given by the well known formula, see for example \cite{yakovlev},
\begin{equation}
\frac{R}{n}=2n\Bigl(\frac{ 2E_{pk}}{3\mu}\Bigr)^{1/2} \frac{S(E_{pk})}{T}\, {\rm e}^{-\tau} {\rm e}^{h^j(0)},
\label{rate}
\end{equation}
Here $\tau=(27\pi^2 \mu Z_i^4 \alpha^2/2T)^{1/3}$, $\mu$ is the reduced mass and $E_{pk}=\tau T/3$.   
We consider a typical temperature of $5.8\times 10^8$ K \cite{gupta}.  The reaction rate versus density is collected in Table \ref{tablefour}.  These results assume the $S$ factor from Table \ref{tableone}, the ion screening from Eq. (\ref{h0analytic}), and the number fraction of $^{24}$O is 0.1.  Oxygen will burn at a density where the reaction rate per ion, times the time for a fluid element to be buried to a given density, is one.  It can take of order 1000 years ($3\times 10^{10}$ s), depending of course on the accretion rate, for a fluid element to be buried to these densities.  We conclude from Table \ref{tablefour} that $^{24}$O will burn at a density near $10^{11}$ g/cm$^3$.

The neutron rich $^{24}$O has a lower fusion rate than $^{16}$O because of its larger reduced mass.  However, most of this reduction in rate is compensated for by a much larger $S$ factor.  Our $S$ factor in Table \ref{tableone} is over eight orders of magnitude larger than the $S$ factor for $^{16}$O+$^{16}$O fusion.  As a result the thermonuclear rate Eq. (\ref{rate}) for $^{24}$O is only slightly smaller than the rate for $^{16}$O fusion.  Indeed, if there is a significant enhancement in $^{24}$O+$^{24}$O fusion because of the dynamics of the neutron rich skin, the $^{24}$O rate will be larger than the $^{16}$O rate.

\begin{table}
\caption{Reaction rate for $^{24}$O+$^{24}$O fusion versus density $\rho$.  The coulomb parameter is $\Gamma$, while the ratio of the temperature to the ion plasma frequency is $T/\omega_p$, and the fusion rate per ion is $R/n$.} 
\begin{tabular}{llll}
 $\rho$ (g/cm$^3$) & $\Gamma$ & $T/\omega_p$ & R/n (s$^{-1}$) \\
\toprule
$10^{10}$ & 170.0 & 2.2 & $6.9\times 10^{-23}$\\
$4\times 10^{10}$ & 269.9 & 1.1 & $4.5\times 10^{-17}$\\
$10^{11}$ & 366.3 & 0.68 & $1.2\times 10^{-11}$\\
$2\times 10^{11}$ & 461.5 & 0.49 & $2.4\times 10^{-6}$\\
\end{tabular} 
\label{tablefour}
\end{table}

Table \ref{tablefour} also lists the ratio of temperature to ion plasma frequency, Eq. (\ref{omegap}).  Strictly speaking the thermonuclear fusion rate, Eq. (\ref{rate}), is only valid for $T>\omega_p$.  We see that the rates in Table \ref{tablefour} involve an extrapolation of Eq. (\ref{rate}) to $T$ slightly below $\omega_p$.  At these temperatures, there will be some quantum corrections to Eq. (\ref{rate}).  However, the thermally enhanced pycnonuclear fusion rates in \cite{yakovlev} suggest that Eq. (\ref{rate}) is not wildly wrong at these temperatures.  Therefore, quantum corrections should not change our conclusions that $^{24}$O will burn at a density near $10^{11}$ g/cm$^3$.  

We also consider $^{28}$Ne+$^{28}$Ne fusion since $^{28}$Ne is the next heavier nucleus in our rp process ash, after $^{24}$O.  However, because of the larger charge $Z=10$, we find that $^{28}$Ne will not burn until higher densities where the plasma frequency is much larger than the temperature.  Therefore $^{28}$Ne will not burn via thermonuclear fusion.  Instead, it will burn via pycnonuclear or thermally enhanced pycnonuclear fusion.  Using the pycnonuclear rates in ref. \cite{yakovlev} along with our $S$ factor from Table \ref{tableone}, we estimate that $^{28}$Ne will burn at densities near $10^{12}$ g/cm$^3$ (for temperatures near $5.8\times 10^8$K).

The fusion of $^{24}$O releases 0.52 MeV per nucleon while $^{28}$Ne fusion releases 0.64 MeV per nucleon.  These energies are larger than the total heating from all of the electron captures considered by Gupta et al. \cite{gupta}.  Therefore, these and other related fusion reactions may be an important source of heat in the crust of accreting neutron stars.

\section{Summary and Conclusions}
\label{summary}
Fusion reactions in the crust of an accreting neutron star are an important source of heat, and the depth at which these reactions occur is important for determining the temperature profile of the star.  Fusion reactions depend strongly on the nuclear charge $Z$.  Nuclei with $Z\le 6$ can fuse at low densities in a liquid ocean.  However, nuclei with $Z=8$ or 10 may not burn until higher densities where the crust is solid and electron capture has made the nuclei neutron rich.

In Section \ref{crosssections} we calculated the $S$ factor for fusion reactions of neutron rich nuclei including $^{24}$O + $^{24}$O and $^{28}$Ne + $^{28}$Ne.  We used a simple barrier penetration model.  We find that $S$ for $^{24}$O+$^{24}$O is over eight orders of magnitude larger than that for $^{16}$O+$^{16}$O.  The $S$ factor could be further enhanced by dynamical effects involving the neutron rich skin of $^{24}$O.  For example, the skins of the two nuclei could deform to form a neck that would reduce the Coulomb barrier.  This possible enhancement in $S$ should be studied in the laboratory with neutron rich radioactive beams.

In Section \ref{MD} we modeled the structure of the crust with molecular dynamics simulations.  We find that the crust of accreting neutron stars may contain micro-crystals or regions of phase separation.  Nevertheless, the screening factors that we determined for the enhancement of the rate of thermonuclear reactions are insensitive to these features.  Finally, we calculated in Section \ref{reaction rates} the rate of thermonuclear $^{24}$O + $^{24}$O fusion and find that $^{24}$O should burn at densities near $10^{11}$ g/cm$^3$.  This is a lower density than some previous estimates.  The 0.52 MeV per nucleon energy released may be important for the temperature profile of the star.  In future work, we will use our molecular dynamics results to study other properties of the crust such as its thermal conductivity.  In addition, we will use these MD results to calculate pycnonuclear reaction rates for the fusion of $^{28}$Ne and other heavier nuclei.         

\section{Acknowledgments}
We thank Ed Brown, Andrew Cumming, Barry Davids, and Romualdo De Souza for helpful discussions and acknowledge the hospitality of the Institute for Nuclear Theory where this work was started.  This work was supported in part by DOE grant DE-FG02-87ER40365 and by Shared University Research grants from IBM, Inc. to Indiana University.

%%%%%%%%%%%%%%%%%%%%%%%%%%%%%%%%%%%%%%%%%%%%%%%%%%%%%%%%%%%%%%%%%
\vfill\eject


\begin{thebibliography}{99} 
%\bibitem{wdwarf} B. M. S. Hansen and J. Liebert, Ann. Rev. Astr. and Astrophysics {\bf 41} (2003) 465.
% M. H. Montgomery, E. W. Klumpe, D. E. Winget, and M. A. Wood, ApJ {\bf 525} %(1999) 482.
\bibitem{rpash} H. Schatz et al., PRL {\bf 86} (2001) 3471.
\bibitem{gupta} S. Gupta, E. F. Brown, H. Schatz, P. Moller, and K-L. Kratz, ApJ {\bf 662} (2007) 1188.
\bibitem{pycno} E. E. Salpeter and H. M. Van Horn, ApJ {\bf 155} (1969) 183.  S. Schramm and S. E. Koonin, ApJ {\bf 365} (1990) 296; erratum {\bf 377} (1991) 343.
\bibitem{superbursts} A. Cumming and L. Bildsten, ApJ {\bf 559} (2001) L127.  
\bibitem{superbursts2} T. E. Strohmayer and E. F. Brown, ApJ {\bf 566} (2002) 1045.
\bibitem{superignition}  A. Cumming, J. Macbeth, J. J. M. in 't Zand and D. Page, ApJ. {\bf 646} (2006) 429.
\bibitem{Wijnands} R.~Wijnands et al., astro-ph/0405089.
\bibitem{cackett} E. M. Cackett et al., MNRAS {\bf 372} (2006) 479. 
\bibitem{rutledge} R. E. Rutledge et al., ApJ. {\bf 580} (2002) 413.
\bibitem{shternin} P. S. Shternin, D. G. Yakovlev, P. Haensel, and A. Y. Potekhin, MNRAS {\bf 382} (2007) L43.
\bibitem{haensel} P. Haensel and J. L. Zdunik, A\& A {\bf 229} (1990) 117 ;  {\bf 404} (2003) L33.
\bibitem{brownprivate} E. F. Brown, private communication.
\bibitem{haensel2007} P. Haensel and J. L. Zdunik, Astronomy + Astrophysics {\bf 480} (2008) 459.
\bibitem{subbarrier} V. I. Zagrebaev, V. V. Samarin, and W. Greiner, PRC {\bf 75} (2007) 035809.
\bibitem{ocpscreening} A. I. Chugunov, H. E. DeWitt, and D. G. Yakovlev, PRD {\bf 76} (2007) 025028.
\bibitem{bimscreening} S. Ogata, S. Ichimaru, and H. M. Van Horn, ApJ. {\bf 417} (1993) 265.
\bibitem{horowitz} C. J. Horowitz, D. K. Berry, and E. F. Brown, PRE {\bf 75} (2007) 066101.
\bibitem{rpash2} S. E. Woosley, A. Hager, A. Cumming, R. D. Hoffman, J. Pruet, T. Rauscher, J. L. Fisker, H. Schatz, B. A. Brown, and M. Wiescher, ApJ Supp. {\bf 151} (2004) 75.
\bibitem{carbon} H. W. Becker et al., Z. Phys. {\bf A303} (1981) 305. 
\bibitem{oxygen} J. Thomas et al., PRC {\bf 31} (1985) 1980. 
\bibitem{bp} A. B. Balantekin, S. Koonin, and J. Negele, PRC {\bf 28} (1983) 1565.
\bibitem{jiang} C. L. Jiang, K. E. Rehm, B. B. Back, and R. V. F. Janssens, PRC {\bf 75} (2007) 015803.
\bibitem{sp} L. R. Gasques et al., PRC {\bf 69} (2004) 034603.  L. C. Chamon et al., PRC {\bf 66} (2002) 014610.
\bibitem{NL3} G. A. Lalazissis, J. Konig, and P. Ring, PRC {\bf 55} (1997) 540.
\bibitem{gasques07} L. R. Gasques et al., PRC {\bf 76} (2007) 045802.
\bibitem{enhancement} L. C. Vaz and J. M. Alexander, PRC {\bf 18} (1978) 2152.
\bibitem{gasques} L. R. Gasques et al., PRC {\bf 72} (2005) 025806.
\bibitem{yakovlev} D. G. Yakovlev, L. R. Gasques, M. Beard, M. Wiescher, and A. V. Afanasjev, PRC {\bf 74} (2006) 035803.
\bibitem{mcocp}H. E. Dewitt, W. L. Slattery, and J. Yang in ``Strongly Coupled Plasmas'', eds. H. M. Van Horn and S. Ichimaru, Univ. of Rochester Press 1993, p425.
\bibitem{jancovici} B. Jancovici, J. Stat. Phys. {\bf 17} (1977) 357.
\bibitem{OCPh} Y. Rosenfeld, PRE {\bf 53} (1996) 2000.
%\bibitem{jones2005} P. B. Jones, Phys. Rev. D {\bf 72} (2005) 083006.
%\bibitem{haskell} B. Haskell, D. I. Jones, and N. Andersson, MNRAS {\bf 373} (2006) 1423.
%\bibitem{peng} F. Peng, E. F. Brown, and J. W. Truran, ApJ {\bf 654} (2007) 1022.
%\bibitem{jones88} P. B. Jones, MNRAS {\bf 233} (1988) 875.
%\bibitem{jones} P. B. Jones, PRL {\bf 83} (1999) 3589.
%\bibitem{potekhin} A. Y. Potekhin and G. Chabrier, Phys. Rev. E {\bf 62} (2000) 8554.  
%\bibitem{ocp} H. DeWitt et al., Contrib. Plasma Phys. {\bf 41} (2001) 251.
%\bibitem{hamaguchi} S. Hamaguchi et al., Phys. Rev. E {\bf 56} (1997) 4671.
%\bibitem{cophasediagram} L. Segretain and G. Chabrier, Astron. Astrophys. {\bf 271} (1993) L13.
%S. Ichimaru, H. Iyetomi, and S. Ogata, ApJ {\bf 334} (1988) L17.
\bibitem{verlet} L. Verlet, Phys. Rev. {\bf 159}, 98 (1967).
F. Ercolessi, A Molecular Dynamics Primer, available from http://www.sissa.it/furio/ (1997).
\bibitem{mdgrape} J. Makino, T. Fukushige, M. Koga, and E. Koutsofias, 
                  in Proceeding of SC2000, Dallas, 2000.
%\bibitem{superobserve} Cornelisse et al., A\& A {\bf 357} (2000) L21.
%\bibitem{sbo2} T. E. Strohmayer, and L. Bildsten in ``Compact Stellar X-ray Sources'', ed. W. Lewin and M. van der Klis, Cambridge University Press (2004) 113.              
%\bibitem{piro} A. L. Piro and L. Bildsten, ApJ. {\bf 629} (2005) 438.
%\bibitem{thermalcond} N. Itoh and Y. Kohyama, ApJ {\bf 404} (1993) 268.



\end{thebibliography}
\end{document}